\documentclass[a4paper,11pt]{article}
\usepackage{jheppub}
\usepackage[utf8]{inputenc}
\usepackage{bbold}
\usepackage{bbm}
\usepackage{bm}
\usepackage{caption}
\usepackage{subcaption}
\usepackage{hyperref}
\usepackage{mathtools,slashed}
\usepackage{mathtools}
\usepackage{xcolor}
\usepackage{appendix}
\usepackage{amsmath,amssymb}

\usepackage{graphicx}
\usepackage{cleveref}
\usepackage{enumitem}
\usepackage[T1]{fontenc}

\newcommand{\te}{\textemdash}

\newcommand{\bracket}[1]{\left\langle #1 \right\rangle}

\hypersetup{
    colorlinks=true,
    linkcolor=blue,
    citecolor=cyan,
    urlcolor=blue,
    }
    
\title{Bubble-wall speed with loop corrections}

\author[a]{Andrii Dashko,}
\author[b]{Andreas Ekstedt}

\affiliation[a]{%
	Deutsches Elektronen-Synchrotron DESY,\\
	Notkestr.~85, 22607 Hamburg, Germany
}

\affiliation[b]{%
    Department of Physics and Astronomy\\
    Uppsala University\\
    Box 516, SE-751 20 Uppsala,
    Sweden
	}

\emailAdd{andrii.dashko@desy.de, andreas.ekstedt@physics.uu.se}

\abstract{In this paper, we investigate the dynamics of the nucleating scalar field during the first-order phase transitions by incorporating one-loop corrections of classical fluctuations. We assume that a high-temperature expansion is valid\te where the mass of the scalar field is significantly smaller than the temperature\te so that we can treat the bubble-wall dynamics in a regime where quantum fluctuations can be integrated out. We present a systematic framework for calculating classical loop corrections to the wall speed; contrast our results with traditional methods based on the derivative expansion; show that the latent heat can differ from the effective-potential result; and discuss general hydrodynamic corrections. Finally, we show an application of the presented framework for a simple scalar field model, finding that the one-loop improvement decreases the wall speed and that an effective-potential approximation underestimates full one-loop corrections by about a factor of two.}


\preprint{DESY-24-169}

\begin{document}

\maketitle
\flushbottom

\newpage

\section{Introduction}\label{sec:Intro}
A gravitational-wave concert from a first-order phase transition could be just the tune needed to explain the observed matter-antimatter asymmetry through Baryogenesis \cite{Cline:2006ts}. Moreover,  gravitational waves provide a window into the Early Universe when it was only nanoseconds old, as well as a glimpse of the Higgs field dynamics during the breaking of Electroweak symmetry \cite{Caprini:2018mtu}. To wit, the phase transition marked the breaking of the Electroweak symmetry of the Standard Model by the Higgs field, which in turn provided masses to particles such as electrons. And things would be even more exciting if this phase transition was violent. While continuous phase transitions, like superheated water, are rather innocuous, the same can not be said about first-order phase transitions. Not only is latent heat released as the transition transpires, but as in boiling water, bubbles of the new phase can form and grow. In the Higgs case, the interior of these bubbles leaves the Electroweak symmetry broken, while the outside conserves it. The release of energy as these bubbles grow can even cause waves in the hot plasma of the Early Universe\te sound waves violent enough to create a symphony of gravitational waves \cite{Kamionkowski:1993fg,Hindmarsh:2017gnf,Caprini:2019egz,Caprini:2007xq}.

This is why the LISA experiment~\cite{amaroseoane2017laserinterferometerspaceantenna} has the community in zest; as one of its core scientific objectives is to look for gravitational waves from cosmological phase transitions.

Yet the Standard Model does not predict a first-order Electroweak transition: New physics is needed \cite{Kajantie:1995kf,Kajantie:1996mn, Laine:2000xu,Aoki:1999fi}. Needless to say, with decades of particle phenomenology, there is no lack of proposals. Nor lack of free parameters. Thus, large parameter scans searching for viable phase transitions are mandated. To get something useful, the methods used for these scans must both be (reasonably) accurate and fast. While the need for speed leaves perturbative calculations as the only option, said calculations are notoriously thorny when it comes to phase transitions~\cite{Athron:2023xlk,Lofgren:2023sep,Gould:2021oba,Gould:2023ovu}.

A lot of the problems arise from the numerous length scales. For example, the nucleation and growth of bubbles during a first-order phase transition are classical processes that occur at length scales of order $L\gg T^{-1}$. Quantum fluctuations, on the other hand, occur on much smaller scales of order $L\sim T^{-1}$. Methods that treat classical and quantum processes at the same time are invariably problematic; due to the presence of large logarithms if nothing else. That is why methods that exploit this large separation of scales are better suited. Like, for example, high-temperature effective theories where predictions are done in a classical theory wherein quantum corrections are incorporated in effective couplings \cite{Farakos:1994kx}.

Such effective methods have already been used to calculate equilibrium quantities such as latent heats and critical temperatures \cite{Farakos:1994kx,Farakos:1995dn,Gould:2023jbz,Kierkla:2023von,Lewicki:2024xan,Niemi:2024vzw}; and near equilibrium quantities, such as bubble-nucleation rates \cite{Gould:2021ccf,Lofgren:2021ogg}. 

In this paper we focus on calculating the terminal velocity of the bubbles: The wall speed. This is a hard quantity to calculate, and the calculations usually require both dealing with kinetic theory and hydrodynamics~\cite{Moore:1995si,Ai:2023see,DeCurtis:2022hlx,Laurent:2022jrs,Ai:2021kak,Dorsch:2021nje,Konstandin:2010dm,Bodeker:2009qy,Bodeker:2017cim}. It is common in these calculations to approximate the scalar potential\te that enters the calculations\te as the effective potential. This procedure, while reasonable as a first approximation, is ad-hoc. And one invariably has to deal with taking roots of negative squared masses that appear. These strange terms come from classical fluctuations, and so can in principle be handled numerically. For example, in the case of nucleation rates such fluctuations give rise to functional determinants.

In this paper we systematically add one-loop classical fluctuations to the wall speed calculations; thus extending partial derivative-expansion results~\cite{Moore:2000wx}. The result is analogous to incorporating fluctuations around the bounce for nucleation-rate determination \cite{Callan:1977pt}. We both provide a general framework for these calculations and give specific examples.

The remainder of the work is structured as follows. In \cref{sec:calculation} we discuss the dynamics of the bubbles during first-order phase transitions and show how the wall speed can be determined. In \cref{sec:prop} we
calculate the Higgs propagator in an inhomogeneous background (with some further details provided in \cref{app:Gkcalc,app:Hydro}). We also comment on multifield propagators in \cref{app:MultiField} and the derivative expansion of the propagator in \cref{app:Derivative}. Next, in \cref{sec:oneloopwall}, we connect the determined Green's functions to bubble-wall velocity. The application of these results for the specific models is discussed in \cref{sec:application}, where a comparison with the effective potential approximation is illustrated.
Finally, we give a summary of the results in \cref{sec:conclusions}.

\section{Background and setup}\label{sec:calculation}
In this paper, we describe the motion of bubble walls during first-order phase transitions. We will assume throughout this paper that a high-temperature expansion is applicable. That is, the typical mass of the Higgs field is smaller than the temperature: $m_H\sim g^2 T\ll T$. Here $g$ represents a generic coupling, which is assumed to be small: $g \ll 1$.  Given these assumptions, the width of the bubble wall is of order $m_H^{-1}\gg T^{-1}$. As such, we can integrate out the quantum fluctuations that occur on length scales $L\sim T^{-1}$. This is then reflected in the fact that our effective mass parameters and couplings implicitly depend on the temperature through matching relations~\cite{Farakos:1994kx,Farakos:1995dn},. Schematically, these matching relations are of the form
\begin{align}
    m^2\rightarrow m^2+ \# T^2, \quad g^2\rightarrow g^2+ \# g^4\log T, \; \ldots
\end{align}

\subsection{Classical fluctuations}
Another effect is that quantum fluctuations\te or equivalently plasma particles\te will resist the motion of any external scalar field or fluid flow. In general such dissipative effects are rather complicated. So in this paper, we assume that the collision rates are large enough for a Chapman-Enskog expansion to be applicable\footnote{We should mention that this assumption is by no means necessary for this paper, but the resulting formulas are simpler.}, and that we can incorporate the dissipation effectively in a friction-parameter $\eta$:
\begin{align}
    \ddot{\phi}=\Vec{\nabla}^2\phi-H'(\phi, \theta)-\eta u\cdot\partial \phi,
\end{align}
where $u^\mu$ is the fluid velocity. $H(\phi, \theta)$ is a Hamiltonian, which includes both the scalar tree-level potential $V(\phi)$ and interactions with other fields $H_\text{int}(\phi, \theta)$; interacting fields are collectively denoted here as $\theta$.

We can conveniently split the scalar field into the background field configuration (bubble profile) $\overline{\phi}$ and fluctuation $\hat{\phi}$: $\phi = \overline{\phi} + \hat{\phi}$, where, on average, the fluctuations vanish $\bigl\langle \hat{ \phi} \bigr\rangle=0$ and the thermally averaged background configuration satisfies $\left<\overline{ \phi} \right>=\overline{\phi}$.

The remaining physics is purely classical\te in the statistical mechanics sense\te and fluctuations occur on length scales $L\sim (g^2 T)^{-1}$. An elegant way to incorporate such fluctuations is to introduce noise via the fluctuations-dissipation theorem~\cite{Bodeker:1999ey,Arnold:1998cy}:
\begin{align}\label{eq:ScalarEOM}
    \ddot{\phi}=\Vec{\nabla}^2\phi-H'(\phi, \theta)-\eta u\cdot\partial \phi+\zeta(\Vec{x},t),
\end{align}
where we require the noise, $\zeta(\vec{x},t)$, to satisfy
\begin{align}
    \bracket{\zeta(x)\zeta(x')}=2 \eta T \delta^{(4)}(x-x'), \quad \bracket{\zeta(x)}=0.
\end{align}

A validation check of this setup is to consider the free ($V'=m^2\phi, H_\text{int}=0$) scalar-field propagator in the plasma frame ($u^0=1$) where one easily finds (for late times $t\gg \eta^{-1}$)
\begin{align}
\left\langle \hat{\phi}(\Vec{k},t)\hat{\phi}(\Vec{k'},t)\right\rangle=\delta^{(3)}(\Vec{k}+\Vec{k'})\frac{T}{\Vec{k}^2+m^2}.
\end{align}

Note that the Langevin setup is by no means necessary. For example, explicit averaging over initial states~\cite{Aarts_1997} is also possible. Both setups are straightforward to simulate numerically, and can as such be compared against the result of the next sections.

\subsection{Finding the wall speed}
Given equation \eqref{eq:ScalarEOM} we can find the wall speed by working in the wall frame where $\dot{\phi}=0$. To be specific, we choose the wall to be in the $z$ direction\footnote{In the main body of the text we consider the wall speed to be constant and so omit any hydrodynamic effects. The general result is given in \cref{app:Hydro}. }. Then we can, to the leading order, omit the noise and solve for the scalar field by enforcing that it reaches the true, and respectively the false, vacuum at $z=\pm \infty$:
\begin{align}
\label{eq:eomLO}
 & \textbf{eom}_{\text{LO}}=   -\overline{\phi}''_0(z)+V'(\overline{\phi})+\omega_0 \overline{\phi}'_0(z)=0,
 \\& \lim_{z\rightarrow \infty} \phi_0(z)= \phi_\text{TV}, \quad \lim_{z\rightarrow -\infty} \phi_0(z)= \phi_\text{FV},
\end{align}
where the friction term comes with the Lorentz-factor $\omega_0\equiv  \eta \gamma_w v_w $ after the boost to the frame of the moving bubble.

This equation can be solved numerically by shooting. This then provides a value for $\omega_0$, $v_w$ and a bubble profile $\overline{\phi}_0 (z)$ for the given $\eta$ and $V'(\phi)$.

\subsection{Next-to-leading order corrections} \label{subsec:nlocor}
Given a leading-order value for the wall speed and a bubble profile, we can then find corrections in a perturbative manner. Later in this paper, we will add one-loop corrections explicitly, but for illustrative purposes imagine that our equations of motion are augmented by some small term $g \Omega(z)$. Then, we can expand our original equation of motion around the leading-order field
\begin{align}
     \overline{\phi}_0(z)\rightarrow \overline{\phi}_0(z)+ g \overline{\phi}_1(z), \quad \omega_0\rightarrow\omega_0+ g \omega_1.
\end{align}
Consequently the $\mathcal{O}(g)$ corrections obey the equation
\begin{align}
\label{eq:eom1}
 \textbf{eom}_{\text{NLO}}= \left(-\partial_z^2+\omega_0 \partial_z+V''(\overline{\phi}_0(z)\right)\overline{\phi}_1(z)+\Omega(\overline{\phi}_0(z),z)+\omega_1 \overline{\phi}'_0(z).
\end{align}
This is just an inhomogeneous ODE that can be solved via the Green function of the operator $\left(-\partial_z^2+\omega_0 \partial_z+V''(\overline{\phi}_0(z)\right)$. This means that we can directly find $\overline{\phi}_1(z)$:
\begin{align}
\label{eq:phi1}
\overline{\phi}_1(z)=-\int d z' G_0(z,z')\left[\Omega(\overline{\phi}_0(z'),z')+\omega_1 \overline{\phi}'_0(z') \right].
\end{align}

However, note that a solution does not generally exist due to the zero mode $\overline{\phi}'_0(z)$. As such we need to use the Fredholm alternative, which specifies that a solution only exists if
\begin{align}\label{eq:NLO_WallSpeed}
\omega_1=-\frac{\int dz' \overline{\phi}'_0(z') e^{-\omega_0 z'}\Omega(\overline{\phi}_0(z'),z')}{\int dz' (\overline{\phi}'_0(z'))^2e^{-\omega_0 z'}},
\end{align}
where we have made the operator self-adjoint by the shift $\overline{\phi}_1(z)\rightarrow \overline{\phi}_1(z) e^{z \omega_0/2}$.

As such, the next-to-leading-order wall speed can be found via \cref{eq:NLO_WallSpeed} without having to explicitly solve the equation of motion \cref{eq:eom1}\footnote{For the example given in the \cref{sec:application}, we numerically solved \cref{eq:eom1} to verify that \cref{eq:phi1,eq:NLO_WallSpeed} hold.}.

\section{Propagators in an inhomogeneous background}\label{sec:prop}
 
In this section, we work out the Higgs field propagator in the inhomogeneous background bubble-wall profile\footnote{See \cite{Ho:2024kzr} for an alternative way to find the Green's function.}; the propagator is simply the correlator of the field's fluctuations $\hat{\phi}$. 
For other particles\te Goldstones, other scalars, and vectors\te the result follows in exactly the same manner. 

We assume that each field is thermalized in a Langevin setup. However, we stress that this assumption is not required; and the result without damping~\cite{Aarts_1997}, for example, follows after setting the damping to zero in the end.

In this section, we treat the wall speed\te and friction\te as a constant, thus neglecting hydrodynamical effects. This is just to save ink, and we discuss the full result with non-constant friction coefficients and hydrodynamics backreactions in \cref{app:Hydro}.

\subsection{Higgs propagator}

Given the assumptions listed above, the Higgs propagator is the Green's function of the operator
\begin{align}\label{eq:eomHiggs}
\widehat{L}_H \hat{\phi}=\left(-\partial_z^2-\vec{\nabla}^2_\perp+\omega_0 \partial_z+V''(\overline{\phi}_0(z)\right)\hat{\phi},
\end{align}
where $\omega_0=\eta \gamma_w v_w\left.\right\vert_\text{LO}$, evaluated through \cref{eq:eomLO}. As it stands, this operator is not self-adjoint, so it is useful to perform the change of basis $\hat{\phi}\rightarrow e^{-\omega_0/2 z}\hat{\phi}$:
\begin{equation}
 \label{eq:eomHiggsAdj}
\widehat{L}_H \rightarrow\left(-\partial_z^2-\vec{\nabla}^2_\perp+W(z)\right), \quad W(z)\footnote{For other particles, the only modification of the equation will be in the function $W(z).$}\equiv V''(\overline{\phi}_0(z))+\frac{\omega_0^2}{4}.
\end{equation} 
The Higgs propagator $G(\vec{x},\vec{y}) = \left< \hat{\phi}(\vec{x},t)\hat{\phi}(\vec{y},t)\right>$ satisfies the equation
\begin{align}
\widehat{L}_H G(\vec{x},\vec{y})=\delta^{3}(\vec{x}-\vec{y}).
\end{align}
We can further use the cylindrical symmetry of $\widehat{L}_H$ to write
\begin{align}
G(\vec{x},\vec{y})=\int dk \sum_m e^{i m(\phi-\phi')} J_m(k \rho)J_m(k \rho') G_k(z,z'),
\end{align}
where $G_k(z,z')$ obeys the equation
\begin{align}\label{eq:GkEquation}
(\partial_z^2-k^2-W(z)) G_k(z,z')=-2 \delta(z-z').
\end{align}
We now have but to find a one-dimensional Green's function for each value of $k$ by solving \cref{eq:GkEquation}. The details of solving this equation can be found in \cref{app:Gkcalc}. 

The single point propagator, the quantity necessary to calculate one-loop corrections to the bubble (see \cref{sec:oneloopwall}), is then simply given by 
\begin{align}
   G(\vec{x},\vec{x})=\frac{1}{4\pi} \int_0^\infty dk k  G_k(z,z).
\end{align}

\subsection{Getting the full propagator}
In practice, one can only solve equation \cref{eq:GkEquation} for so many $k$, and, moreover, handling UV-divergences numerically is troublesome, hence, we split the integral as follows:

\begin{equation}\label{eq:PropSol}
    \begin{split}
        G(\vec{x},\vec{x})=&\frac{1}{4\pi}\underbrace{\int_0^{\infty}dk k G_k^\text{free}(z,z) }_\text{finite in dimReg}
        \\+& \frac{1}{4\pi} \underbrace{\int_0^{1}dk k \left[G_k(z,z)-G_k^\text{free}(z,z) \right]}_\text{finite} + \frac{1}{4\pi} \underbrace{\int_1^{k_\text{max}}dk k \left[G_k(z,z)-G_k^\text{WKB}(z,z) \right]}_\text{finite}
        \\+& \frac{1}{4\pi} \underbrace{\int_{k_\text{max}}^\infty dk k \left[G_k(z,z)-G_k^\text{WKB}(z,z) \right]}_{\approx 0}+\frac{1}{4\pi} \underbrace{\int_{1}^{\infty} dk k \left[G_k^\text{WKB}(z,z)-G_k^\text{free}(z,z) \right]}_\text{finite}
    \end{split}
\end{equation}
where $G_k^\text{free}$ is a free propagator of a scalar particle with the mass $m^2 = W(\infty)$ and for the first integral in \cref{eq:PropSol} the dimensional regularization was employed for UV regularization, leading to effective-potential-like result $-\frac{W(\infty)^{1/2}}{4\pi}$. 

The WKB approximation (see \cref{app:WKB}), which is applicable for large $k$ values, is introduced to accelerate convergence. The total integral, as it should be, does not depend on the arbitrary cut-off $k_{\text{max}}$.

\section{One-loop corrections for a bubble}\label{sec:oneloopwall}
After determining the propagators, as described in the previous section, for all particle content in the given model, one has to include these corrections into the background field equation of motion. For a more detailed discussion of the determination of multifield propagators, refer to \cref{app:MultiField}.

Starting from \cref{eq:ScalarEOM}, we can write for the background field

\begin{align}
    \ddot{\overline{\phi}}=\Vec{\nabla}^2\overline{\phi}-\eta \dot{\overline{\phi}} - V'(\overline{\phi})-\frac{1}{2}G_{H} V'''(\overline{\phi})-\frac{1}{2}G_{\theta_i \theta_j} \left. \frac{\partial^2 H'_\text{int}(\overline{\phi}, \theta)}{\partial \theta_i \partial \theta_j} \right|_{\theta=0} + \dots,
\end{align}
where $\theta$ stands for a set of fluctuating fields interacting with the scalar field $\phi$; $G_{\theta_i \theta_j} = \left< \theta_i \theta_j \right> $\textendash Green's function, correlator of these fluctuating fields. Note here that there could be nonvanishing off-diagonal terms, for example, Goldstones-vectors mixing.

To connect with the previous section's notation, 
\begin{align}
\label{eq:OmegaDef}
\Omega(\overline{\phi}(z),z) =\frac{1}{2}G_{H}(z,z) V'''(\overline{\phi}(z))+\frac{1}{2}G_{\theta_i \theta_j}(z,z) \left. \frac{\partial^2 H'_\text{int}(\overline{\phi}(z), \theta)}{\partial \theta_i \partial \theta_j} \right|_{\theta=0}.
\end{align}
Once the single-point Green's function in the inhomogeneous background (outlined in \cref{sec:prop}) has been determined for all fluctuating fields coupled to the scalar field $\phi$, we apply \cref{eq:NLO_WallSpeed} and \cref{eq:OmegaDef} to compute the one-loop correction to the wall speed.

\section{Applications}\label{sec:application}

In this section, we apply the formalism described in previous chapters to the specific example and show how the results compare to the naive calculations with the derivative expansion\textendash effective potential.

\subsection{A real-scalar model}\label{sec:realscalar}
We proceed by considering a real singlet scalar field in 3 dimensions with the given tree-level potential
\begin{align}\label{eq:potential}
V(\phi)=\frac{1}{2}y\phi^2-\frac{1}{6}\phi^3+\frac{1}{4}x \phi^4,
\end{align}
where we have already introduced dimensionless parameters $x, y$ and made a specific normalization of the field. This potential could for example arise after the dimensional reduction.
The potential possesses a non-trivial, symmetry-breaking vacuum state $\phi_b=\frac{\sqrt{1-16 x y}+1}{4 x}$ in addition to the symmetric $\phi_s = 0$.

\begin{figure}[ht]
\hspace*{-1cm} 
\centering
\includegraphics[width=1.075\textwidth]{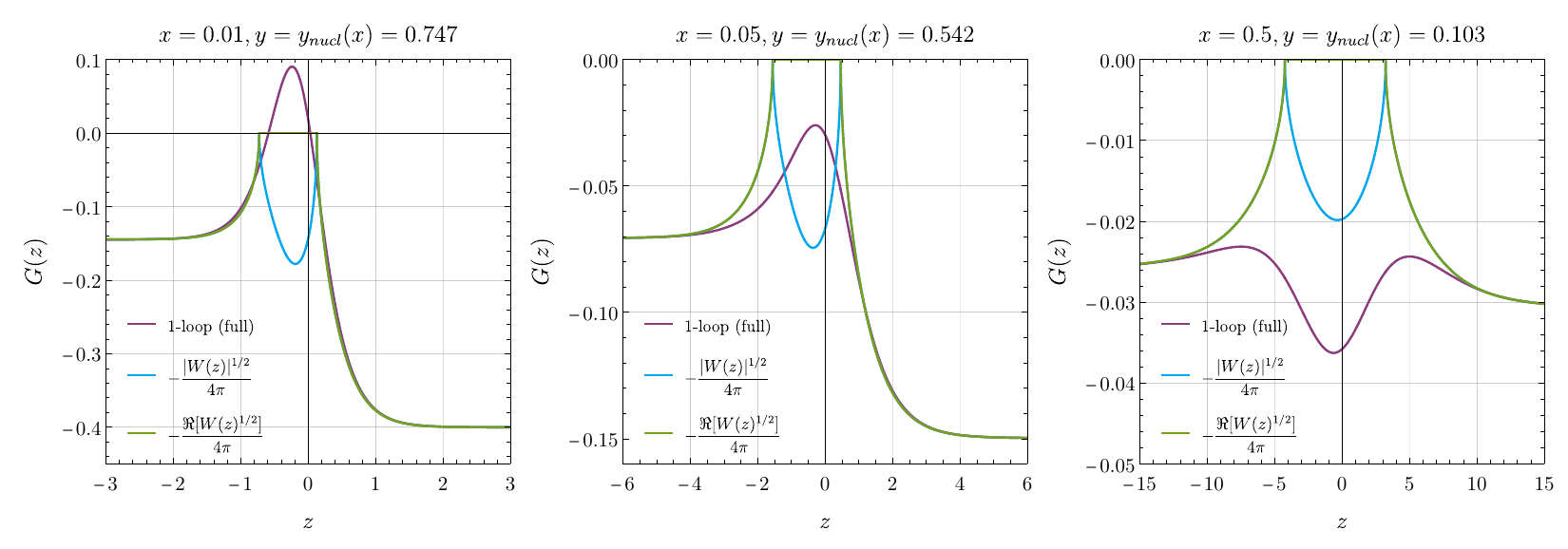}
\caption{One point Higgs Green's function, for $x = 0.01$ (\textit{Left}), $x= 0.05$ (\textit{Center}) and $x = 0.5$ (\textit{Right}) evaluated at the nucleation mass value.}
\label{fig:1}
\end{figure}

\Cref{fig:1} presents the Green's function for the real singlet, computed as outlined in \cref{sec:prop} together with the effective potential approximation. The effective potential approximation corresponds to using the leading order derivative expansion, further explored in \cref{app:Derivative}, resulting in $G(z) = -\frac{\sqrt{W(z)}}{4\pi}$; as $W(z)$ may become negative, there are two options to get real-valued Green's function: taking a real part of the square root $\Re[\sqrt{W(z)}]$ or an absolute value of $W(z)$.

All plots presented are evaluated at the nucleation mass parameter $y = y_{nucl}$ for the given coupling $x$. In other words, $S_{b}(x,y = y_{nucl}) \approx 126$, where the bounce is calculated at the leading order, using results from \cite{Ekstedt:2021kyx}. 

\begin{figure}[ht]
\centering
\hspace*{-1cm} 
\includegraphics[width=1.075\textwidth]{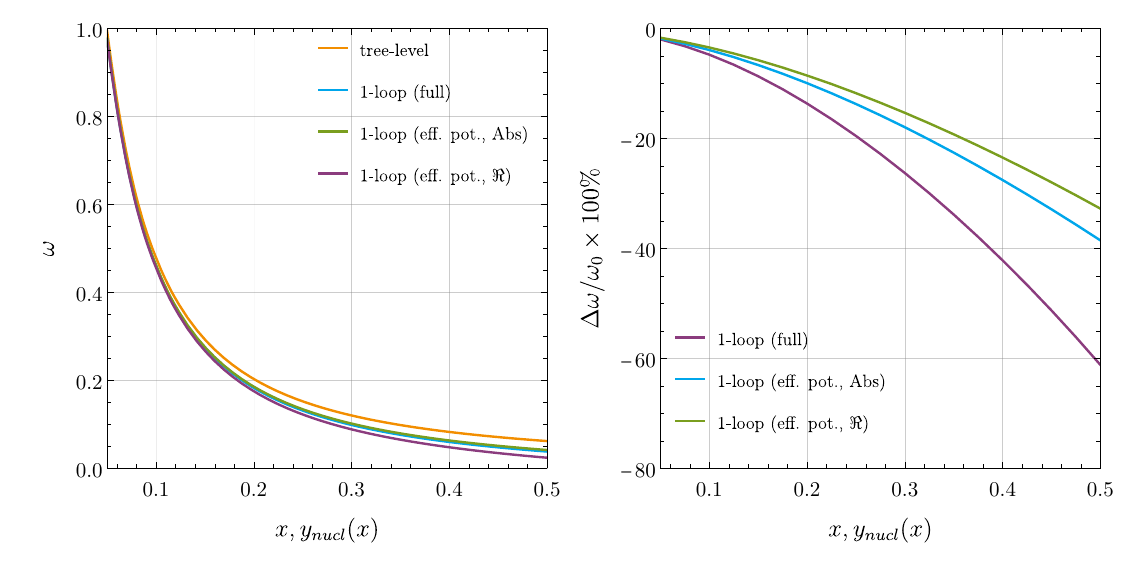}
\caption{Velocity factor $\omega = \eta v_w \gamma_v$ (\textit{Left}) and relative correction to the velocity factor evaluated at tree-level (\textit{Right}). Negative corrections indicate the fact that loop corrections decrease the wall speed. For the effective potential approximation, both choices for the square root are shown.}
\label{fig:2}
\end{figure}

We present the determined according to \cref{eq:NLO_WallSpeed} wall velocity for the given potential \cref{eq:potential} in \cref{fig:2}, where we explicitly taken $\Omega(\phi(z),z) = \frac{1}{2}G_H(z)V'''(\phi(z))$.  As expected, the corrections get bigger for stronger coupling, in the end reaching the region of failure of the perturbative expansion, at most at $ x \approx 0.24 $, at which point the perturbative expansion for the effective action breaks down \cite{Ekstedt:2022ceo}.
The full one-loop correction to the wall speed is about two times larger than the correction resulting from using the effective potential approximation for the whole parameter range in the model explored; both slow down the bubble.

\subsection{Radiative barriers}\label{sec:radbarrier}
In the context of this paper, a radiative barrier arises when (classical) one-loop corrections overpower the tree-level potential to generate a barrier. As an example consider an additional scalar\footnote{The argument for vector-bosons is identical.} $\psi$:

\begin{align}
    H(\phi, \psi) = V(\phi) + \frac{1}{2} g_\psi^2 \phi^2 \psi^2, \quad V(\phi) = \frac{1}{2}m^2\phi^2 +\frac{1}{4}\lambda \phi^4
\end{align}
A barrier can then be generated if the field-dependent mass of $\psi$ dominates the Higgs mass:
\begin{align}
    g_\psi^2 \phi^2 \gg V''(\phi) \sim  m^2.
\end{align}
In this case, the "heavy" scalar $\psi$ can be integrated out, and a barrier can be generated:
\begin{align}
    V(\phi) \rightarrow V(\phi)-\frac{1}{12 \pi}(g_\psi^2 \phi^2)^{3/2}.
\end{align}
Crucially this construction is \textbf{not} valid when $\phi \approx 0$. Consequently, if we wish to find the wall speed consistently we have to modify our procedure. 

In essence, we want to accommodate a barrier when $\phi \gg 0$, while simultaneously capturing the right dynamics of $\psi$ around $\phi\approx 0$. The easiest way to accomplish this is to add and subtract the effective-potential contribution from $\psi$\te that is $-g_\psi^2 \overline{\phi}(z)\frac{1}{4\pi}\left(W_\psi(\overline{\phi})\right)^{1/2}$\te to the equation of motion of the background field while adding the full contribution from the $\psi$ propagator.

The one-loop correction then becomes
\begin{align}
\label{eq:OmegaDef0}
\Omega(\overline{\phi}(z),z) =\frac{1}{2}G_{H}(z,z) V'''_\text{LO}(\overline{\phi}(z))+g_\psi^2 \overline{\phi}(z) \left (G_{\psi \psi}(z,z)- \left[- \frac{1}{4\pi}\left(W_\psi(\overline{\phi}(z))\right)^{1/2} \right]\right),
\end{align}
and the leading-order potential, which enters the LO equation of motion \cref{eq:eomLO}, becomes
\begin{align}
  V_\text{LO}(\phi) = V(\phi)  -  \frac{1}{12\pi}\left(W_\psi(\phi)\right)^{3/2}, \quad W_\psi = g_\psi^2 \phi^2+\frac{\eta_\psi^2\gamma_w^2 v_w^2}{4}.
\end{align}

\begin{figure}[ht]
\hspace*{-1cm} 
\centering
\includegraphics[width=1.075\textwidth]{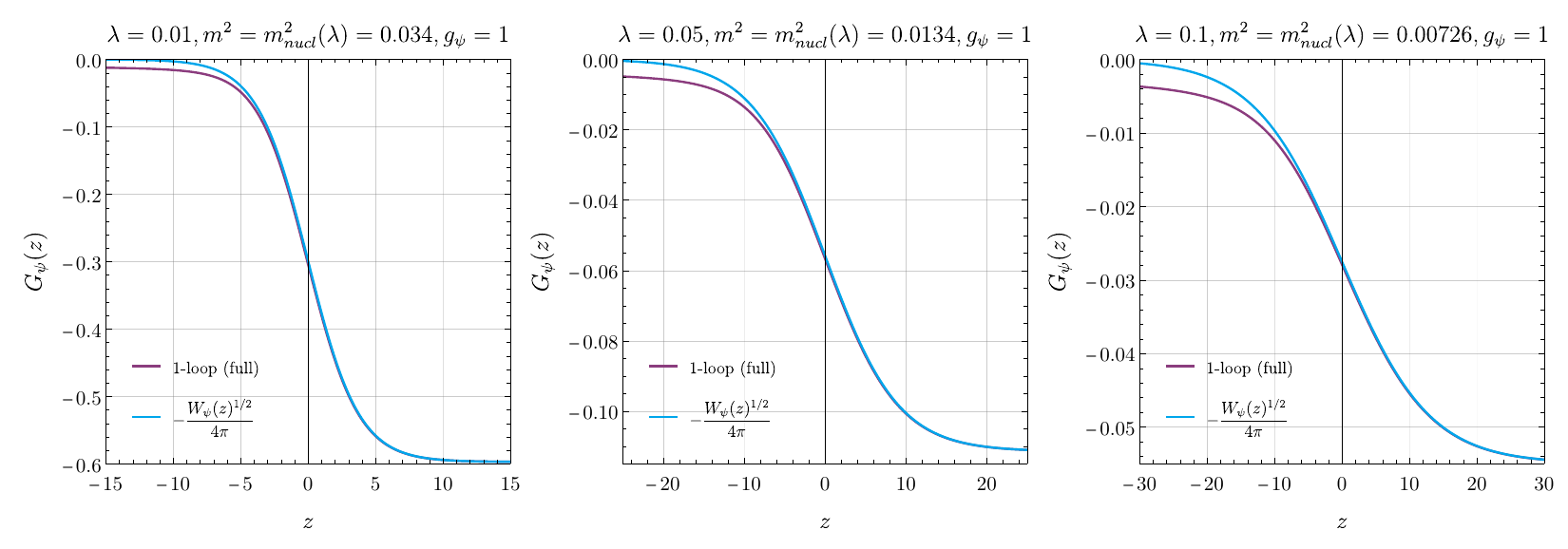}
\caption{"Heavy" scalar Green's function $G_{\psi\psi}(z)$ for $\lambda = 0.01$, $g_\psi = 1$ (\textit{Left}), $\lambda = 0.05$, $g_\psi = 1$ (\textit{Center}) and $\lambda = 0.1$, $g_\psi = 1$ (\textit{Right}) evaluated at the nucleation mass value. The $\psi$-field damping is set to zero $\eta_\psi=0$.}

\label{fig:3}
\end{figure}

We want to stress two things here.  First, we have included the $\eta_\psi$ term in $W_\psi$ because the $\psi$ field is in general damped. Second, there is no reason to suspect that $G_{\psi \psi}(z,z)$ approaches the normal effective-potential result when  $z\rightarrow -\infty$. See for example \cref{fig:3}. Both of these effects can significantly reduce the wall speed by lowering the effective latent heat. For example, the transition becomes second-order when $(\eta_\psi \gamma_w v_w)^2 \sim m^2/\lambda$.
This implies that radiative barriers can not give rise to runaway bubbles ($v_w \rightarrow 1$). Though it should be stressed that our assumptions break down for large enough wall speeds, so this statement should be taken with a grain of salt.

\subsection{Consequences for hydrodynamics}
In the previous section, we saw an example when the $\lim_{z\rightarrow \pm \infty}$ behaviour of the scalar equation of motion differed from the effective-potential prediction. This is also important for hydrodynamical considerations~\cite{Ai:2023see,DeCurtis:2022hlx,Laurent:2022jrs,Ai:2021kak,Dorsch:2021nje,Konstandin:2010dm,Bodeker:2009qy,Bodeker:2017cim}; for example by giving a different latent heat.

\section{Conclusions}\label{sec:conclusions}

In this work, we have assessed the effect of loop corrections on the equation of motion of the nucleating scalar field during the first-order phase transition, focusing on the impact on the speed of the bubble wall.

The work is split into two main parts: first, the correlator of the nucleating scalar field fluctuations with an inhomogeneous background was calculated, and then the impact of these fluctuations on the bubble profile and bubble-wall velocity was determined. To our knowledge, this is the first systematic study of the effects of one-loop fluctuations of the scalar field on the bubble-wall speed. 

To illustrate the calculations, we have further considered an example of a nucleating real scalar field in three dimensions. It was shown that one-loop corrections decrease the wall speed, reaching tens of percent correction to the tree-level velocity, giving the highest corrections at the border of perturbativity. The effective-potential approximation\textemdash leading-order derivative expansion term\textemdash underestimates the decrease of the wall speed by approximately a factor of two. Higher terms in the derivative expansion do not improve the situation. Thus signaling the failure of the validity of the aforementioned expansion for the bubble.

The findings presented in this paper may be used in further studies in more realistic scenarios, by including larger particle content and adding hydrodynamical effects. The framework for handling more complex cases, such as non-constant fluid velocity, multifield propagators, and hydrodynamical backreactions, has been presented in the appendices.

\section*{Acknowledgments}
A.D acknowledges support from the
Deutsche Forschungsgemeinschaft (DFG, German Research Foundation) under Germany's Excellence Strategy – EXC 2121 ``Quantum Universe'' - 390833306 and the Deutsche Forschungsgemeinschaft 
(DFG, German Research Foundation) - 491245950. The work of A.E.\ has been supported by the Swedish Research Council, project number VR:$2021$-$00363$.

\appendix

\section{Finding \texorpdfstring{$G_k(z,z')$}{Gk(z,z')}} \label{app:Gkcalc}
To find the propagator we need to find the homogeneous solutions of equation \eqref{eq:GkEquation}. The asymptotic behavior of the solutions is
\begin{align}
&z\rightarrow -\infty: \ G_k^{<}\sim \exp\left[z \sqrt{k^2+W(-\infty)} \right], \ G_k^{>}\sim \exp\left[-z \sqrt{k^2+W(-\infty)} \right],
\\& z\rightarrow \infty: \ G_k^{<}\sim \exp\left[-z \sqrt{k^2+W(\infty)} \right], \ G_k^{>}\sim \exp\left[z \sqrt{k^2+W(\infty)} \right].
\end{align}
So to facilitate a numerical solution it is auspicious to define
\begin{align}
\label{eq:GkAnz}
&G_k^{<}=A^{<}(z')T_k(z) \Psi^{<}(z), \; G_k^{>}=A^{>}(z')H_k(z) \Psi^{>}(z),
\\& \Psi^{</>}(z)=\exp\left[\pm \int_{0}^z \sqrt{k^2+W(s)+F(s)}ds \right].
\end{align}
for $z<z'$ and $z>z'$ respectively.
The full solutions can now be found by solving the following equations numerically:
\begin{equation}
\label{eq:TkRes}
    \begin{aligned}
    &T''_k + 2 \sqrt{k^2 + W(z) + F(z)} \, T'_k + \frac{W'(z) + F'(z)}{2 \sqrt{k^2 + W(z) + F(z)}} \, T_k + F(z) \, T_k = 0 \\
    & \text{with} \ T'_k(-\infty) = 0 \ \text{and} \ T_k(-\infty) = 1,
    \end{aligned}
\end{equation}
\begin{equation}
\label{eq:HkRes}
    \begin{aligned}
    &H''_k - 2 \sqrt{k^2 + W(z) + F(z)} \, H'_k + \frac{W'(z) + F'(z)}{2 \sqrt{k^2 + W(z) + F(z)}} \, H_k + F(z) \, H_k = 0 \\
    & \text{with} \ H'_k(\infty) = 0 \ \text{and} \ H_k(\infty) = 1.
    \end{aligned}
\end{equation}
Here we have introduced an auxiliary function $F(z)$ such that $F(z)+W(z)\geq 0$ for all $z$; this is just to simplify the numerical evaluations, and the final result is, of course, independent of $F(z)$.

Finally, given $T_k(z)$ and $H_k(z)$, the full propagator is given by\footnote{This follows after using the standard jumping conditions at $z=z'$.}
\begin{equation}
\label{eq:GkRes}
    \begin{split}
G_k(z,z')=\left[\sqrt{k^2+W(\infty)}T_k(\infty) \right]^{-1}&\left[ \theta(z-z')T_k(z')\Psi^{<}(z')H_k(z)\Psi^{>}(z) \right.
\\&\left. +\theta(z'-z)T_k(z)\Psi^{<}(z)H_k(z')\Psi^{>}(z') \right]
    \end{split}
\end{equation}

\subsection{Handling zero-modes}
Contrary to other particles, for Higgs and Goldstone propagators hit an additional obstacle due to the presence of zero-modes. Take, for example, the Higgs case, where $W(z)= V''(\overline{\phi}_0(z))+\frac{\omega_0^2}{4}$. The zero mode is then $\overline{\phi}'_0(z)e^{-z \omega_0/2}$; the presence of zero modes is an artifact of the broken symmetry of the equation of motion by the background solution. A normal eigenfunction expansion then implies that
\begin{align}
&G_k(z,z')=\frac{2}{k^2} \frac{\phi'_0(z)e^{-z \omega_0/2}\overline{\phi}'_0(z')e^{-z' \omega_0/2}}{N_\phi}+\ldots,
\\& N_\phi=\int ds \overline{\phi}'_0(s)^2e^{-s \omega_0}
\end{align}
The zero-mode term vanishes in the integral over $k$ if we use dimensional regularization, but small $k$ values are still troublesome numerically. To deal with this snag we subtract 0 from \eqref{eq:PropSol}:
\begin{align}
0 = -\frac{2}{N_\phi}\overline{\phi}'_0(z)^2 e^{-z\omega_0} \int dk k^{-1} =-\frac{2}{N_\phi}\overline{\phi}'_0(z)^2 e^{-z\omega_0} \left[\int_0^{k_\text{max}}+\int_{k_\text{max}}^\infty  \right]dk k^{-1}.
\end{align}
We then explicitly add the first term to equation \eqref{eq:PropSol} for each $k$, while the second term effectively  gives $\frac{2}{N_\phi}\overline{\phi}'_0(z)^2 e^{-z\omega_0}\log k_\text{max}$. 

The procedure is the same for Goldstone modes; for example, for a complex scalar where $W(z)= \overline{\phi}_0(z)^{-1}V'(\overline{\phi}_0(z))+\frac{\omega_0^2}{4}$ the zero mode is $e^{- z\omega_0/2}\overline{\phi}_0(z)$, and the procedure is identical to the Higgs case.

\subsection{The WKB approximation}\label{app:WKB}

In case $k^2 \gg W(z)$ the $z$-dependence of $W(z)$ is just a minor consideration, which allows us to solve \cref{eq:GkEquation} within the WKB approximation.
For $G_k^{</>}$, expanding for large $k$, we find
\begin{align}
    \log G_k^{</>} \approx \pm\left[ k z\pm \frac{1}{2k}\int ds W(s)\right]+\log\left|1-\frac{1}{4k^2}W(z)\right|.
\end{align}
Similar expansion can be made with the functions that appeared in our ansatz \cref{eq:GkAnz}:
\begin{align}
\Psi^{</>}(z)= \exp\left[\pm\int_{0}^z \sqrt{k^2+W(s)}ds \right]\approx  \exp\left[\pm k z \pm \frac{1}{2k}\int_0^zds W(s) \right].
\end{align}
Lastly, after imposing the boundary conditions, we find the asymptotic expansion for $T_k$ and $H_k$:
\begin{align}
\label{eq:TkHkWKB}
&T_k\approx 1-\frac{W(z)-W(-\infty)}{4 k^2},
\\& H_k\approx 1-\frac{W(z)-W(\infty)}{4 k^2}.
\end{align}

\section{Non-constant fluid velocity}\label{app:Hydro}

\subsection{Without one-loop hydrodynamic backreactions}
Consider again the operator
\begin{align}\label{eq:eomHiggsH}
\widehat{L}_H \hat{\phi}=\left(-\partial_z^2-\vec{\nabla}^2_\perp+\omega_0(z) \partial_z+V''(\overline{\phi}_0(z))\right)\hat{\phi},
\end{align}
now with a z-dependent $\omega_0$, as occurs if we solve the full hydrodynamical equations. A first approximation is to incorporate $\omega_0(z)$ without adding the $\hat{\phi}$ dependence of the wall speed and temperature.

In this case, we can make this operator self-adjoint with the shift $\hat{\phi}\rightarrow e^{-\frac{1}{2}\int_0^z \omega_0(s) ds}\hat{\phi}$:
\begin{align}
   \widehat{L}_H \rightarrow \left(-\partial_z^2-\vec{\nabla}^2_\perp+V''(\overline{\phi}_0(z)) +\frac{\omega_0^2}{4}-\frac{1}{2}\omega_0'(z)\right).
\end{align}
The results in Section \ref{sec:prop} are the same after defining $W(z)=V''(\overline{\phi}_0(z)) +\frac{\omega_0^2}{4}-\frac{1}{2}\omega_0'(z)$. Note also that the new Higgs zero-mode is $e^{-\frac{1}{2}\int_0^z \omega_0(z) ds} \overline{\phi}'_0(z)$.

Finally, to find corrections to the leading-order wall speed we simply use $e^{-\int_0^z \omega_0(z) ds} \overline{\phi}'_0(z)$ in place of $e^{-z \omega_0(z) } \overline{\phi}'_0(z)$ in equation \eqref{eq:NLO_WallSpeed}.

\subsection{Propagators with general equations of motion}

Consider the general equation of motion
\begin{align}
  &  \phi''(z)+V'(\phi,\theta)+F(\phi,\phi',\theta)=0,
\end{align}   
where for example $\theta$ might represent the fluid velocity or the temperature. After we expand around a leading-order solution, $\phi=\overline{\phi}+\hat{\phi}$, we then get
\begin{align}
  &  \hat{\phi}''(z)+ \delta_{\phi}\left[V'(\phi,\theta)+F(\phi,\phi',\theta)\right]_{\phi=\overline{\phi}} \hat{\phi}(z)=0.
\end{align}  
Here $F$ might be a local friction term as in \cref{sec:oneloopwall}, but $F$ could also come from the solution of a Boltzmann equation~\cite{Moore:1995si}, and would then, in general, be a non-local function of $\phi$.
In taking the $\phi$ derivative we have to take into account the $\phi$ dependence of $\theta$; for example a field-dependent damping coefficient or the temperature. All in all, this ensures that $\overline{\phi}'$ is a zero-mode of the above operator.

The propagator can then be found as usual if we can evaluate $\frac{\partial \theta}{\partial \phi}$.

\subsection{Adding one-loop hydrodynamic backreactions}
As an example of the previous subsection, consider adding one-loop hydrodynamic corrections\footnote{We here neglect the bulk and shear viscosity. If these are present we would also expect velocity and temperature fluctuations.}. First, we need the hydrodynamical equations in the wall frame:
\begin{align}
&w v_w \gamma_w^2 =\text{const},
\\&\frac{1}{2}(\phi')^2-V+w v_w^2 \gamma_w^2=\text{const},
\end{align}
where the enthalpy is of the form $w=\frac{4}{3}a T^4- T \partial_T V$. If we omit the second-term we can then solve for $v_w$ and $T$ to the first order in $\hat{\phi}$:
\begin{align}
    \delta v=\underbrace{\frac{1}{a T^4 \gamma_w^2 v_w}}_{=a_w}\left[V'(\overline{\phi}_0) \hat{\phi}-\overline{\phi}_0' \hat{\phi}'\right], \quad \delta T=-\underbrace{\frac{v_w^2+1}{4 a T^3 v_w^2}}_{=b_w}\left[V'(\overline{\phi}_0) \hat{\phi}-\overline{\phi}_0' \hat{\phi}'\right]
\end{align}
Adding these corrections to the Higgs equation of motion we find
\begin{equation}
\label{eq:eomHiggsH2}
    \begin{split}
        \widehat{L}_H \hat{\phi}=&\left(-\partial_z^2-\vec{\nabla}^2_\perp+\omega_0(z) \partial_z+V''(\overline{\phi}_0(z))\right)\hat{\phi}+\left[a_w \partial_{v_w} \omega_0 \overline{\phi}'_0+b_w \partial_T V'(\overline{\phi}_0(z)) \right]\hat{\phi}
        \\&-\left[a_w \partial_{v_w} \omega_0 \overline{\phi}'_0+b_w \partial_T V'(\overline{\phi}_0(z)) \right]\hat{\phi}'.    
    \end{split}
\end{equation}
This operator can now be made self-adjoint as in the previous sub-section, and propagators for $T(z), v_w(z),$ and $\hat{\phi}(z)$ can be found using the results in Section \ref{sec:calculation}.

Note that $\phi$ propagators will also enter the conservation of the energy-momentum tensor in various places.

\section{Multifield propagators}\label{app:MultiField}
The procedure to find multifield propagators numerically readily follows from the single-field case that we have delineated in \cref{sec:prop} of this article. Consider for example an equation of the form
\begin{align}
\left[-\vec{\nabla}^2\delta_{\theta_i \theta_j}+\omega_{\theta_i \theta_j}(z)\partial_z +M^2_{\theta_i \theta_j}(z) \right]G_{\theta_i \theta_j}(\vec{x}-\vec{y})=\delta_{\theta_i \theta_j} \delta^{3}(\vec{x}-\vec{y}),
\end{align}
where $\theta_i, \theta_j$ runs over all relevant fields; $\omega_{\theta_i \theta_j}(z)$ and $M^2(z)_{\theta_i \theta_j}$ are friction- and mass-matrices respectively.

The linear z-derivative terms can be removed by the shift
\begin{align}
    G_{\theta_i \theta_j}\rightarrow \left\{e^{\frac{1}{2} \int_0^z \omega }G e^{-\frac{1}{2} \int_0^{z'} \omega }\right\}_{\theta_i \theta_j}.
\end{align}
This brings the equation of motion to the form:
\begin{align}
    \left[-\vec{\nabla}^2\delta_{\theta_i \theta_j}+W_{\theta_i \theta_j}(z) \right]G_{\theta_i \theta_j}(\vec{x}-\vec{y})=\delta_{\theta_i \theta_j} \delta^{3}(\vec{x}-\vec{y}),
\end{align}
where
\begin{align}
W_{\theta_i \theta_j}(z)=e^{-\frac{1}{2} \int_0^z \omega }\left[M^2_{\theta_i \theta_j}(z)+\frac{1}{4}\omega^2_{\theta_i \theta_j}-\frac{1}{2}\partial_z \omega_{\theta_i \theta_j} \right]e^{\frac{1}{2} \int_0^z \omega }.
\end{align}

To find the propagators one can follow the same steps as before. Numerically, it is the simplest to first rotate to an eigenbasis of $W_{\theta_i \theta_j}(-\infty)$, evolve the equations to $z=0$, rotate to an eigenbasis of $W_{\theta_i \theta_j}(\infty)$, and then evolve the system to $z=\infty$. We note that an additional complication arises if mixing with vector fields is present, in which case non-trivial momentum dependence of mass matrix perplexes the calculation.

\section{Derivative expansion}\label{app:Derivative}
An analytic approximation to the propagator (derivative expansion) can be obtained under the assumption that the field-dependent mass is large: $\partial_z^2 \ll M^2(z)$. To derive a derivative expansion recall the equation of motion for the propagator:
\begin{align}
    \left[-\partial^2_z -\vec{\nabla}_\perp^2 +W(z)\right]G(\vec{x},\vec{x}')=\delta^{3}(\vec{x}-\vec{x}').
\end{align}
Then, following \cite{Moss:1985ve}, one can set $\vec{x}'=\vec{x}+\vec{y}$ and solve for $G$ in powers of $\partial_z$. Subsequently, one finds
\begin{equation}
\label{eq:derexpansion}
    \begin{split}
        G(\vec{x},\vec{x})=-&\frac{1}{4\pi}\sqrt{W(z)}+\frac{1}{512 \pi W(z)^{5/2}}\left(\partial_z W(z)\right)^2-\frac{1}{328 \pi W(z)^{3/2}}\partial_z^2 W(z) \\ 
        +& \frac{105}{32768 \pi^2 W(z)^{11/2}}\left(\partial_z W(z)\right)^4-\frac{77}{12288 \pi^2 W(z)^{9/2}}\left(\partial_z W(z)\right)^2 \partial^2_z W(z) \\
        +&\frac{3}{2048 \pi^2 W(z)^{7/2}}\left(\partial^2_z W(z)\right)^2 + \frac{1}{512 \pi^2 W(z)^{7/2}}\partial_z W(z)\partial^3_z W(z) \\ 
        +&\frac{9}{2560 \pi^2 W(z)^{5/2}}\partial^4_z W(z)
        \ldots \: .
    \end{split}
\end{equation}

\begin{figure}[ht]
    \hspace*{-1cm} 
    \centering
    \includegraphics[width=1.075\textwidth]{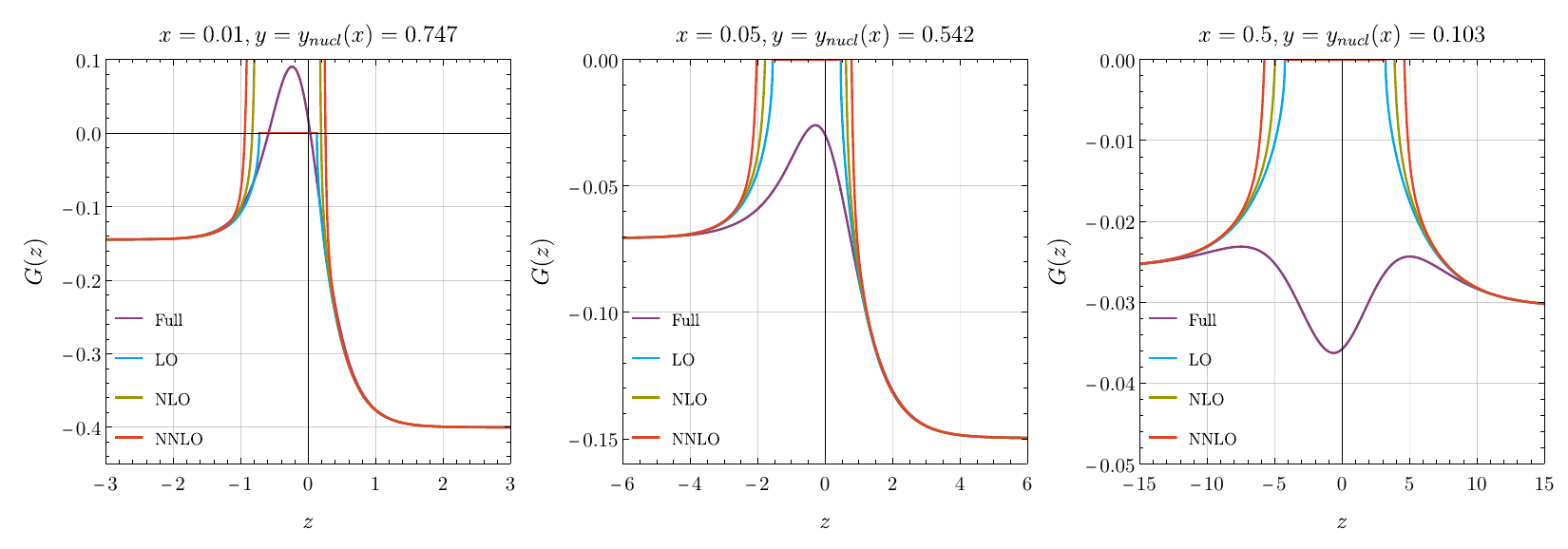}
    \caption{Higher derivative expansion of Higgs Green's function terms together with the full numerically evaluated propagator. A real part of Green's function is taken for the derivative expansion, omitting the spurious imaginary part.}
    \label{fig:4}
\end{figure}

\Cref{fig:4} displays the numerically evaluated Higgs propagator alongside the leading-order (LO), as well as the next-to-leading-order (NLO) and next-to-next-leading-order (NNLO) derivative expansions for the example model discussed in \cref{sec:realscalar}.
The derivative expansion is valid only at large distances from the bubble wall; thus, incorporating higher-order terms would not converge to the complete propagator in the bubble background, which is essential for accurately determining bubble profile and speed corrections.

\newpage

\bibliographystyle{JHEP}

\bibliography{bibl}

\end{document}